\documentclass[prb,twocolumn,showpacs,amsmath,amssymb,superscriptaddress]{revtex4}
\usepackage[dvips]{graphicx}
\usepackage{graphics}
\usepackage{dcolumn}
\usepackage{bm}
\usepackage{amsmath}
\usepackage{amsmath}

\newcommand{\nc}{\newcommand}

\nc{\be}{\begin{equation}} \nc{\ee}{\end{equation}}
\nc{\bea}{\begin{eqnarray}} \nc{\eea}{\end{eqnarray}}
\nc{\bean}{\begin{eqnarray*}} \nc{\eean}{\end{eqnarray*}}

\begin{document}

\title{Control of Josephson current by Aharonov-Casher Phase in a Rashba Ring}
\author{Xin Liu}
\affiliation{ Department of Physics, Texas A\&M University, College
Station, TX 77843-4242, USA}
\author{M. F. Borunda}
\affiliation{ Department of Physics, Texas A\&M University, College
Station, TX 77843-4242, USA}
\author{Xiong-Jun Liu}
\affiliation{ Department of Physics, Texas A\&M University, College
Station, TX 77843-4242, USA}
\author{Jairo Sinova}
\affiliation{ Department of Physics, Texas A\&M University, College
Station, TX 77843-4242, USA}
\affiliation{Institute of Physics ASCR, Cukrovarnick\'a 10, 162 53 Praha 6, Czech
Republic }
\date{\today}
\begin{abstract}
We study the interference effect induced by the Aharonov-Casher  phase on the Josephson current through a semiconducting ring attached to superconducting leads.
Using a 1D model that incorporates spin-orbit coupling in the semiconducting ring, we
calculate the Andreev levels analytically and numerically, and predict oscillations of the Josephson current due to the AC phase.
This result is valid from the point contact limit to the long channel length limit, as defined by the ratio of the junction  length and
the BCS healing length. We show in the long channel length limit that 
the impurity scattering has no effect on the oscillation of the Josephson current, 
in contrast to the case of conductivity oscillations in a spin-orbit coupled ring system attached to normal leads where impurity scattering reduces the amplitude of oscillations.
Our results suggest a new scheme to measure the AC phase with, in principle, higher sensitivity. 
In addition, this effect allows for control of the Josephson current through 
the gate voltage tuned  AC phase. 
\end{abstract}
\maketitle

\section{Introduction}

When a quantum system undergoes cyclic evolution motion in 
parameter space, its wave function acquires a geometric phase which strongly influences the transport property of the system.
The best known example of such phase is the Aharonov-Bohm (AB)
phase which is the relative phase shift between two charged particle
paths enclosing a magnetic flux. In the last few years, the Aharonov-Casher (AC) phase,\cite{A-C} another example of a geometric phase, has
been studied both experimentally \cite{AC experiment1,AC
experiment2} and theoretically \cite{AC,Nikolic,path} in spin-orbit (SO) coupled mesoscopic semiconducting rings. As the charge-spin dual of the AB  effect, the AC phase in semiconductors has
attracted great attention because it can be easily controlled by a gate voltage. So far,
most of the literature has focused on Rashba SO coupled rings attached to
normal metal leads. The AC phase has been confirmed in these
type of systems by observing the conductance oscillation
as a function of the gate voltage that is related to the SO interaction (SOI)
strength. However, due to the large background current, the typical amplitude of
the oscillation is no more than 10\% of the total observed
conductance.\cite{AC experiment2} The AC phase has also been
discussed for magnetic vortices in type-II superconductors and Josephson junction array.\cite{AC vor1,AC
vor2,AC vor3,AC vor4} Recent studies in the effect of
SOI on Josephson
current\cite{2002jso,2006jra,2007jqd,2008jra,2008jso} has focused on  heterostructures comprised of a two-dimensional electron
gas (2DEG) or a quantum dot placed between two superconducting leads. 
A natural and interesting step is to combine the AC and Josephson effects to explore new regimes in semiconductor-superconductor structures, 
which has not been explored to our knowledge.

\begin{figure}[ht]
\centering
\includegraphics[width=0.8\columnwidth]{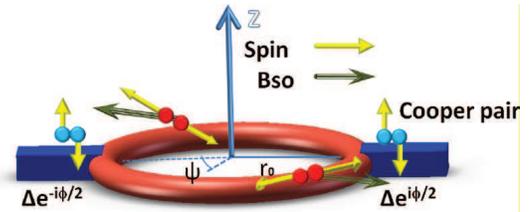}
\caption{Two superconductor leads are attached at the points
$\psi=0$ and $\psi=\pi$ separately on the ring, Josephson
current can be obtained in this junction.}
\label{ring}
\end{figure}

In this article, we study the AC effect in a Josephson current
traversing a ring with Rashba type SO interaction. The Josephson current in a
superconducting-normal-superconducting (S/N/S) junction mainly originates from the
Andreev reflection,\cite{Andreev1,RMPAndreev} (electrons being
retroreflected to holes) in two Superconducing/Normal (S/N)
interfaces. The excitation spectrum below the superconducting gap
consists of discrete energy levels, called Andreev levels.\cite{Andreev2}
Andreev levels are affected by both the phase and the
transmission coefficient of electrons and holes in the normal
region.\cite{Kulik,Bagwell,prl1991}
 We study here the effects of SOI on the Andreev levels and 
 the corresponding Josephson current through the ring, illustrated in Fig.~\ref{ring}, both analytically and numerically. Our calculations predict  oscillations
of the Josephson current controlled by the AC phase.
This effect can be used as an alternative sensitive way to detect the AC phase.
Because the dephasing electrons will not contribute to the current in a Josephson junction but will
contribute to the current in a normal junction attached to two non-supercoducting leads, 
the observed  amplitude of the current
oscillations due to AC effect in a ring Josephson junction should be much larger than 
the conductance oscillations in the  normal ring junction.

The outline of this paper is as follows. In Sec. II, we introduce
our theoretical model of the Josephson current in a 1-D mesoscopic
SO coupled ring-shaped Josephson junction. In Sec. III, we use the
boundary conditions of a multiple terminal junction and 
present the numerical calculations of the Andreev levels and the Josephson currents in limits not explored 
by the analytical solution. In Sec. IV we present an experimental set-up for the observation 
of the effect and in Sec. V summarize our conclusions. In Appendix A, we
present the details of the sub-gap states of quasiparticles in
superconductor. In appendix B we show the details of calculating the eigenstates and
 the eigenenergies of the ring in the tight binding model.

\section{Analytical discussion of Andreev spectrum due to spin-orbit coupling}

\begin{figure}[ht]
\centering
\includegraphics[width=0.55\columnwidth]{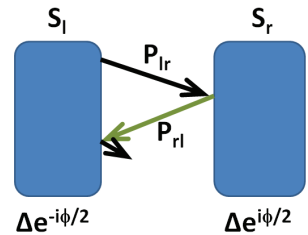}
\caption{A step-like superconducting pair potential is assumed in
the Josephson junction. The right moving particle in the normal
region will come back to its initial state after two  Andreev
reflections. The change of the color of the arrow after each reflection
means an electron is reflected to a hole and a hole is reflected to
an electron.}
\label{fig1}
\end{figure}

The Andreev bound state is a kind of electron-hole bound state in the normal region of the Josephson junction depicted by
Fig.~\ref{fig1}. In the superconductor leads, the excitation spectrum consists of the
positive eigenvalues of the Bogoliubov equation
\begin{eqnarray}\label{bog}
\left(\begin{array}{cc}H(r)&\Delta\\\Delta^*&-H(r)\end{array}\right)\Psi=E\Psi,\\
H(r)=-\frac{\hbar^2}{2m^*}\nabla^2+U(r)-\mu,
\end{eqnarray}
where $H(r)$ is the one-electron Hamiltonian, $\mu$ is the
electrochemical potential, $U(r)$ is the scalar potential, $\Delta$
is the superconducting pair potential and $m^*$
is the effective mass. When the excitation energy is less than the gap
$|\Delta|$, 
the quasiparticles of Eq.~(\ref{bog}) in the normal region will be
 reflected by the pairing potential and form the Andreev bound state.
Unlike the bound states in a usual square
well, the Andreev levels carry  electric current which contributes to
most of the Josephson current.\cite{prl1991} Generally speaking, the bound
states can be described in such a way that if we have a right moving
subgap particle on the left S/N interface Fig.~\ref{fig1}, after gaining a phase described by the
matrix $P_{lr}$ due to the particle propagation from left to right, reflection on the right interface
described by the scattering matrix $S_r$,  a further phase described by the matrix $P_{rl}$ from 
the motion from right to left, and a reflection on the left interface with the scattering matrix
$S_l$, the state should come back to its original state. This is
true only when ${\rm Det}(\mathbf{1}-P_{rl}S_rP_{lr}S_l)=0$, which determines the
quantized Andreev levels.

In zero magnetic field and zero SOI, neglecting Fermi
wavelength mismatch and the barrier on the S/N interface and
assuming that the transmission function of electrons in normal region is
energy independent, these discrete energies are determined
by:\cite{Bagwell}
\begin{eqnarray}\label{ANL}
&&2\arccos(\frac{E}{\Delta})+\gamma \pm
\theta=2\pi n,\gamma=(\frac{L}{\xi_0})(\frac{E}{\Delta}), \nonumber \\
&&\cos(\theta)=
T\cos(\phi)+R\cos\left[(\frac{L-2a}{\xi_0})\frac{E}{\Delta}\right],
\end{eqnarray}
where $\xi_0=\hbar v_F/2\Delta$ is the BCS healing length,
$\phi=\phi_2-\phi_1$ is the superconducting phase differences,
$n=0,\pm 1,\pm 2 \cdots$,   $\theta$ is the additional
phase due to a point impurity potential $V\delta(x)$ in the normal
region, $a$ is the distance between the point impurity and left
interface, and $T$ and $R$ are corresponding to the transmission
probability and reflection probability due to a point impurity when
energies are close to the fermi energy. The term $(EL/\Delta \xi_0)$
is approximately the phase shift acquired from free electrons and
holes propagation in the normal region. Eq.~(\ref{ANL}) demonstrates
that the Andreev level will be determined by both the phase shift and the
transmission function of electrons and holes in the normal region. An important limit
is when the length of the junction is much shorter than the BCS healing length ($L\ll\xi_0$) so that
$\gamma\simeq0$. In this case, the Andreev level takes the
form\cite{prl1991}
\begin{eqnarray}\label{Andreev level}
E=\Delta\cos(\frac{\theta}{2})=\Delta\sqrt{1-T\sin^2(\frac{\phi}{2})}.
\end{eqnarray}

Next we focus on the Josephson current in a one-dimentional (1-D) Rashba SOI ring; this
current can be computed from the superconducting phase dependence of the Andreev bound states.
For our model calculations we assume a
step-like superconducting pair potential $\Delta$, which
is zero in the Rashba ring:
\begin{eqnarray}\label{potential}
\Delta=\{\begin{array}{l} \Delta_0 e^{-i\phi/2}, \ \ \ \ \ \  \mbox{left of the ring}\\
0, \ \ \ \ \ \ \ \ \ \ \ \ \ \ \ \ \ \mbox{in the ring}\\
\Delta_0 e^{i\phi/2}, \ \ \ \ \ \ \mbox{right of the ring}
\end{array}
\end{eqnarray}
where $+(-)\phi/2$ are the superconducting phase in the right(left) leads.

Because the SOI keeps time-reversal symmetry, the
coefficient of the Andreev reflection is almost the same as without
SOI. As a result, Eq.~(\ref{ANL}) is still valid qualitatively for a
ring-like junction with SOI. The 1D Rashba Hamiltonian for a
1-channel ring of radius $r_0$ without magnetic field reads:\cite{ring}
\begin{eqnarray}\label{1Dham}
H_{1D}=\frac{\hbar \omega_0}{2}(-i\frac{\partial}{\partial
\varphi})^2+\frac{\hbar\omega_r}{2}(\cos\varphi \sigma_x+\sin\varphi
\sigma_y)\nonumber \\(-i\frac{\partial}{\partial
\varphi})-i\frac{\hbar\omega_r}{4}(\cos\varphi \sigma_y-\sin\varphi
\sigma_x),
\end{eqnarray}
where $r_0$ is the radius of the ring, $\omega_0=\hbar/(m^*r^2_0)$ with $m^*$ being the effective mass of the electron in the ring,
and $\omega_R=2\alpha/(\hbar r_0)$ with $\alpha$ being the the strength of the spin-orbit coupling. The eigenfunctions of this
Hamiltonian take the form:
\begin{eqnarray}\label{eigen}
&&
\Psi_{e,n}^{\uparrow}(\varphi)=\exp(in_{e}\varphi)\left(\begin{array}{c}\sin(\gamma/2)e^{-i\varphi}\\\cos(\gamma/2)\end{array}\right),\nonumber\\
&&
\Psi_{e,m}^{\downarrow}(\varphi)=\exp(im_{e}\varphi)\left(\begin{array}{c}\cos(\gamma/2)\\-\sin(\gamma/2)e^{i\varphi}\end{array}\right).
\end{eqnarray}
The associated eigenenergies read
\begin{eqnarray}\label{energy}
E_{e,n}&=&\frac{\hbar
\omega_0}{2}[(n_e-\frac{1}{2})^2+(n_e-\frac{1}{2})\sqrt{1+Q_R^2}+\frac{1}{4}]-\mu, \nonumber
\\
E_{e,m}&=&\frac{\hbar
\omega_0}{2}[(m_e+\frac{1}{2})^2-(m_e+\frac{1}{2})\sqrt{1+Q_R^2}+\frac{1}{4}]-\mu, \nonumber \\
\\
n_{e}&=&\frac{1-\sqrt{1+Q_R^2}\pm\sqrt{Q_R^2+4C_e}}{2},\nonumber
\\ m_{e}&=&\frac{\sqrt{1+Q_R^2}-1\pm \sqrt{Q_R^2+4C_e}}{2},
\end{eqnarray} where $n_{e}$ is the quantized angular momentum number of the electron with
spin along $-z$
direction, $m_{e}$ is the quantized angular momentum number of the electron with spin along $z$ direction, $\mu$ is
the chemical potential of the system and $C_e=\frac{2(E+\mu)}{\hbar
\omega_0}$ is a parameter independent of SOI.

Since the hole is the time-reversal state of the electron and SOI keeps time-reversal symmetry, we have $H_h=-H_e$. Therefore,
the hole states are the same as the electron states and the
associated eigenenergies take the similar form:
\begin{eqnarray}\label{energyh}
E_{h,n}&=&\mu-\frac{\hbar
\omega_0}{2}[(n_h-\frac{1}{2})^2+(n_h-\frac{1}{2})\sqrt{1+Q_R^2}+\frac{1}{4}],\nonumber
\\
E_{h,m}&=&\mu-\frac{\hbar
\omega_0}{2}[(m_h+\frac{1}{2})^2-(m_h+\frac{1}{2})\sqrt{1+Q_R^2}+\frac{1}{4}],\nonumber \\
\\
n_{h}&=&\frac{1-\sqrt{1+Q_R^2}\pm \sqrt{Q_R^2+4C_h}}{2},\nonumber\\
m_{h}&=&\frac{\sqrt{1+Q_R^2}-1\pm \sqrt{Q_R^2+4C_h}}{2},
\end{eqnarray}
where $C_h=\frac{2(\mu-E)}{\hbar \omega_0}$.

We next consider the phase difference between to the electron and hole
propagation in the normal region.  The phase difference in the upper part of the ring for the spin-up
and spin-down quasiparticles  after being Andreev
reflected two times take the form
\begin{eqnarray}\label{gamma}
&&\gamma_{n}=(n_{e}-n_{h})\pi=\frac{\sqrt{Q_R^2+4C_e}-\sqrt{Q_R^2+4C_h}}{2}\pi,\nonumber\\
&&\gamma_{m}=(m_{e}-m_{h})\pi=\frac{\sqrt{Q_R^2+4C_e}-\sqrt{Q_R^2+4C_h}}{2}\pi,\nonumber\\
\end{eqnarray}
and are therefore identical. Here
 $\gamma_{n(m)}$ is the phase shift of the quasiparticle with spin along $-z(+z)$ direction.
This identity reflects the fact that there is no
spin splitting of the phase shift in this case and is consistent with
the time-reversal symmetry of SOI. The phase acquired in the
Andreev reflection from the interfaces lead to a different phase shift since the
spin up electrons and spin down holes have different momenta.\cite{2008jso} 
However this phase shift splitting is very
small and will not affect the zeros of the Josephson current. Therefore
we will not consider this splitting in our discussion.

The effect of the Rashba interaction on the transmission function has been
studied previously, with a conductance oscillation due to the AC phase 
confirmed experimentally\cite{AC experiment1,AC
experiment2} and theoretically.\cite{AC,Nikolic,path} Under the assumption of
a perfect coupling between the leads and the ring and neglecting 
backscattering, the transmission function takes the form\cite{AC}
\begin{eqnarray}\label{Tr}
T&=&\frac{1}{2}+\frac{1}{2}\cos[(\sqrt{1+Q_R^2}-1)\pi]\nonumber \\
&=&\cos^2[(\sqrt(1+Q_R^2)\pi)].
\end{eqnarray}

When considering the short junction limit, $L\ll \xi_0$,
the energy $E$ of the Andreev bound state is affected by the normal-state
transmission  $T$ through Eq.~(\ref{Andreev level}). The
Josephson current in the low temperature limit takes the form
\cite{prl1991}
\begin{eqnarray}\label{JC}
I&=&\frac{-2e}{\hbar}\frac{\partial E}{\partial
\phi}=\frac{e\Delta}{2\hbar}\frac{T\sin(\phi)}{\sqrt{1-T\sin^2(\phi/2)}}\nonumber\\
&=&\frac{e\Delta}{2\hbar}\frac{\cos^2[(\sqrt(1+Q_R^2)\pi)]\sin(\phi)}{\sqrt{1-\cos^2[(\sqrt(1+Q_R^2)\pi)]\sin^2(\phi/2)}}.
\end{eqnarray}
Because the transmission probability is affected by the AC phase, an
oscillation of the Josephson current due to SOI should also be observed.
This oscillation will be different from the conductance oscillation
because even in the short junction limit, the current is nonlinearly
dependent on the transmission function. Although this conclusion is
obtained in the short junction limit ($L\ll \xi_0$), since the zeros of the transmission
probability are only dependent on the SOI, a similar oscillatin
 can be expected at any value of $L/\xi_0$. We will show this to be the case in
our numerical calculations.

\section{Numerical calculation of Andreev level and Josephson current}
In this section, we present our numerical investigation on the
relation of the Josephson current to the SOI. The
effetive tight-binding model Hamiltonian of the 1D Rashba ring
is given
as \cite{Nikolic}:
\begin{eqnarray}\label{tbra}
H_{ring}&=&\sum_{j}^{N}2t_0
\hat{c}^{\dag}_{j}\hat{c}_j+\sum_{j}^{N}\sum_{\sigma,\sigma'}[t_{\varphi}^{j,j+1;\sigma,\sigma'}\hat{c}_{j+1}^{\dag}\hat{c}_j+H.C],\nonumber
\\
t_{\varphi}^{j,j+1}&=&-i
\frac{t_{so}}{(r/a)\delta\varphi}(\cos\varphi_{j,j+1}\hat{\sigma}_x+\sin\varphi_{j,j+1}\hat{\sigma}_y)\nonumber\\
&&+\frac{1}{(r/a)^2(\delta\varphi)^2}t_0
\hat{I}_s,
\end{eqnarray}
where $\varphi_j=2\pi(j-1)/N$,
$\varphi_{j,j+1}=(\varphi_j+\varphi_{j+1})/2$,
$\delta\varphi=2\pi/N$, $t_{so}=\alpha/2a$, $r$ is the radius of the ring
and $t_0=\hbar^2/2ma^2$ with $a$ being the lattice constant of the tight-binding model.
The length scale $a$ is not related to the lattice constant of the material but is an artifical length
scale used in modeling the continum Hamiltonian, Eq.~(\ref{1Dham}), and all 
physical results obtained should be independent of this length scale; e.g. the Fermi energy scales 
considered should be near the  bottom of the tight-binding Hamiltonian band, $E_F\ll 4t_0$. The
eigenfunctions of this Hamiltonian are the same as  in Eq.~(\ref{eigen}).
The eigenenergy is obtained as
\begin{eqnarray}\label{tbenergy}
&&E_{en}=2t_0\left[1-\cos[(n-\frac{1}{2})\delta\varphi+\beta]\sqrt{1+(\frac{t_{so}}{t_0})^2}\right]-\mu\nonumber
\\
&&E_{em}=2t_0\left[1-\cos[(m+\frac{1}{2})\delta\varphi-\beta]\sqrt{1+(\frac{t_{so}}{t_0})^2}\right]-\mu\nonumber\\
&&E_{hn}=-E_{en}, E_{hm}=-E_{em},
\end{eqnarray}
where
\begin{eqnarray}\label{tbang}
&&n_{e(h)}=(\pm\lambda_{e(h)}-\beta)/\delta \varphi+1/2,\nonumber \\
&&m_{e(h)}=(\pm\lambda_{e(h)}+\beta)/\delta \varphi-1/2,\nonumber \\
&&\lambda_{e(h)}=\arccos(\frac{1-(\mu+(-)E)/2t_0}{\sqrt{1+(t_{so}/t_0)^2}}).
\end{eqnarray}
Here
$\beta=\arccos(\frac{cos(\delta\varphi/2)}{\sqrt{1+(t_{so}/t_0)^2}})$.
The detailed derivation of the result shown above 
is presented in Appendix B. Defining the AC phase as
$\phi_{AC}\equiv w_{so}\pi\equiv(2\beta/\delta\varphi-1)$ and comparing $n$ and $m$ in the tight
binding model with those in the continuous case, we can find that the term
$w_{so}\pi=(2\beta/\delta\varphi-1)$ is the counterpart of the term $(\sqrt{1+Q_R^2}-1)\pi$ in the
continuous case and they can be shown to be equal in the limit $a\rightarrow 0$. 
 In the following discussion, we will often use the dimensionless parameter $w_{so}$ instead of $t_{so}/t$
 in our numerical calculations and plots as a measure of the SOI strength since the
 AC phase is a monotomic increasing function of it.

If we ignore the disoreder in the ring, the phase shift matrices
$P_{lr}$ and $P_{rl}$ are diagonal matrices and each diagonal
element takes the form $\exp(i(n_{e(h)}-n_{h(e)})\pi)$. The S
matrices $S_l$ and $S_r$ are calculated by solving the boundary
condition at each joint of the superconductor leads and the ring. In a
ring system, the joint of the lead and the ring can be considered as a
3-way junction. The boundary condition of this 3-way junction in the
tight-binding model is obtained from the principle that the wave
function at the joints is also the eigenfunction of the
system.\cite{Mario} This boundary condition in the tight-binding model allows us to
consider more realistic cases such as having different effective mass of
electrons in the superconducting leads and normal region. Once S
matrices $S_l$ and $S_r$ are obtained, and substituting them into the
bound state condition,\cite{semi-sup} 
\begin{eqnarray}\label{adbound}
\mathrm{Det}(\mathbf{1}-P_{rl}S_rP_{lr}S_l)=0,
\end{eqnarray}
we can find the relation among the Andreev level $E$, superconductor phase
difference $\phi$ and the SOI strength. Eq.~(\ref{adbound}) contains all of the Feyman paths and allows us
 to consider the effect of multiple backscattering within the ring. 

In the previous section, to get the analytical result, we made the assumption that the
coupling between the ring and the leads are perfect and backscattering was therefore neglected.
However, this idealization is not a good approximation in 
 general since there is usually a very large Schottky barrier on
the superconductor/semiconductor interface and the backscattering can also
strongly affects the electron transmission function through a ring. All of these factors
are considered in our numerical calculation.

First, let us focus on the effects of the backscattering.  We consider the limit $L\ll \xi_0$. In the tight binding model,
$\xi_0=\hbar v_f/2\Delta=a t_0\sin(k_f a)/\Delta$.  We assume $\Delta=0.001t_0$, $E_f=t_0$ ({\em i.e.} 
one quarter of the band-width), $\xi_0=500\sqrt{3} a$ and 
the perimeter of the ring is $100 a$ which is smaller than the healing coherence length $\xi_0$. We choose $E_f=t_0$ because the 
transmission probability of the electron through the ring oscillates slower in this energy range compared to its zero minimal value and 
 when the energy is close to the bottom of the band. As a result, the electron's behavior in this energy range satisfies the approximation 
 we made in the previous section and more directly confirms our analytical discussion. 
 The parameters corresponding to the experimental system\cite{AC experiment1} will be discussed at the end of this section.

\begin{figure}[ht]
\centering
\includegraphics[width=0.8\columnwidth]{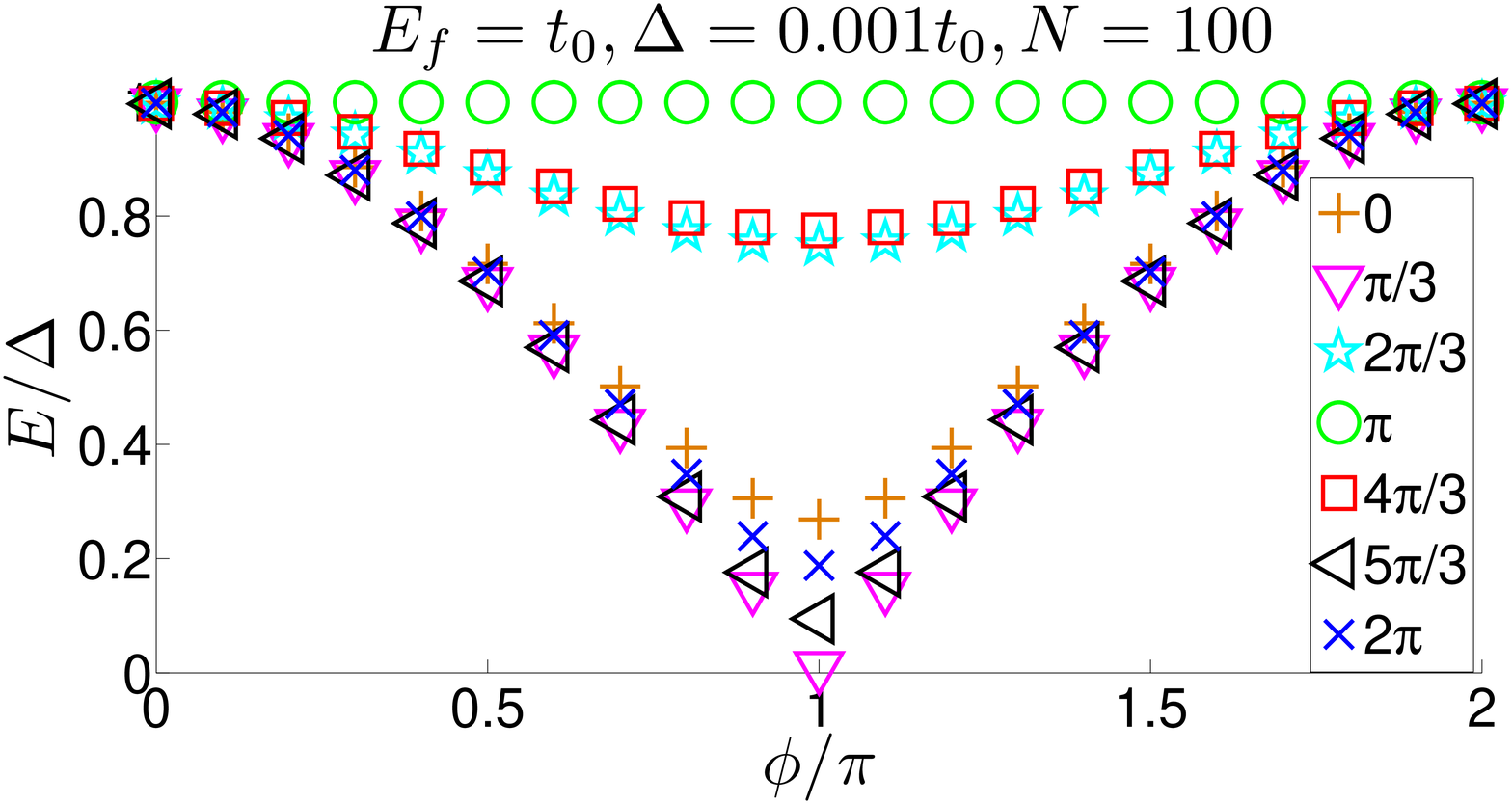}
 \hfill
\includegraphics[width=0.8\columnwidth]{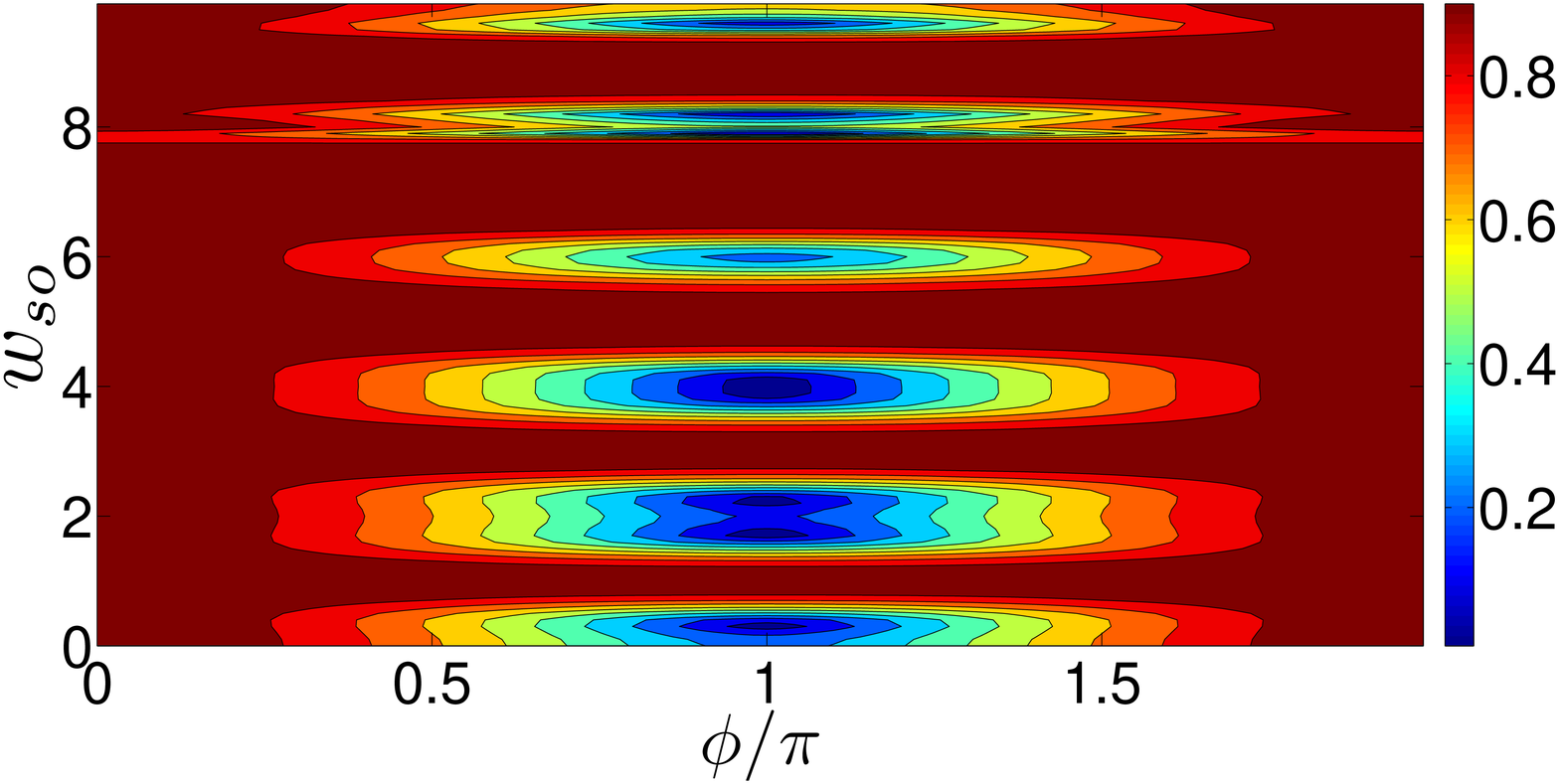}
\caption{Top: Andreev levels vs. superconducting phase difference
 $\phi$ for different SOI strength measured by AC phase($w_{so}\pi$). Bottom: Andreev levels (color plot) vs.
superconducting phase difference,
 $\phi$, and SOI measured by $w_{so}$. 
 Here $N=100$, $E_f=t_0$, $\Delta=0.001t_0$, and $\xi_0=500\sqrt{3} a$.}
 \label{Al1}
\end{figure}
\begin{figure}
\centering
\includegraphics[width=0.9\columnwidth]{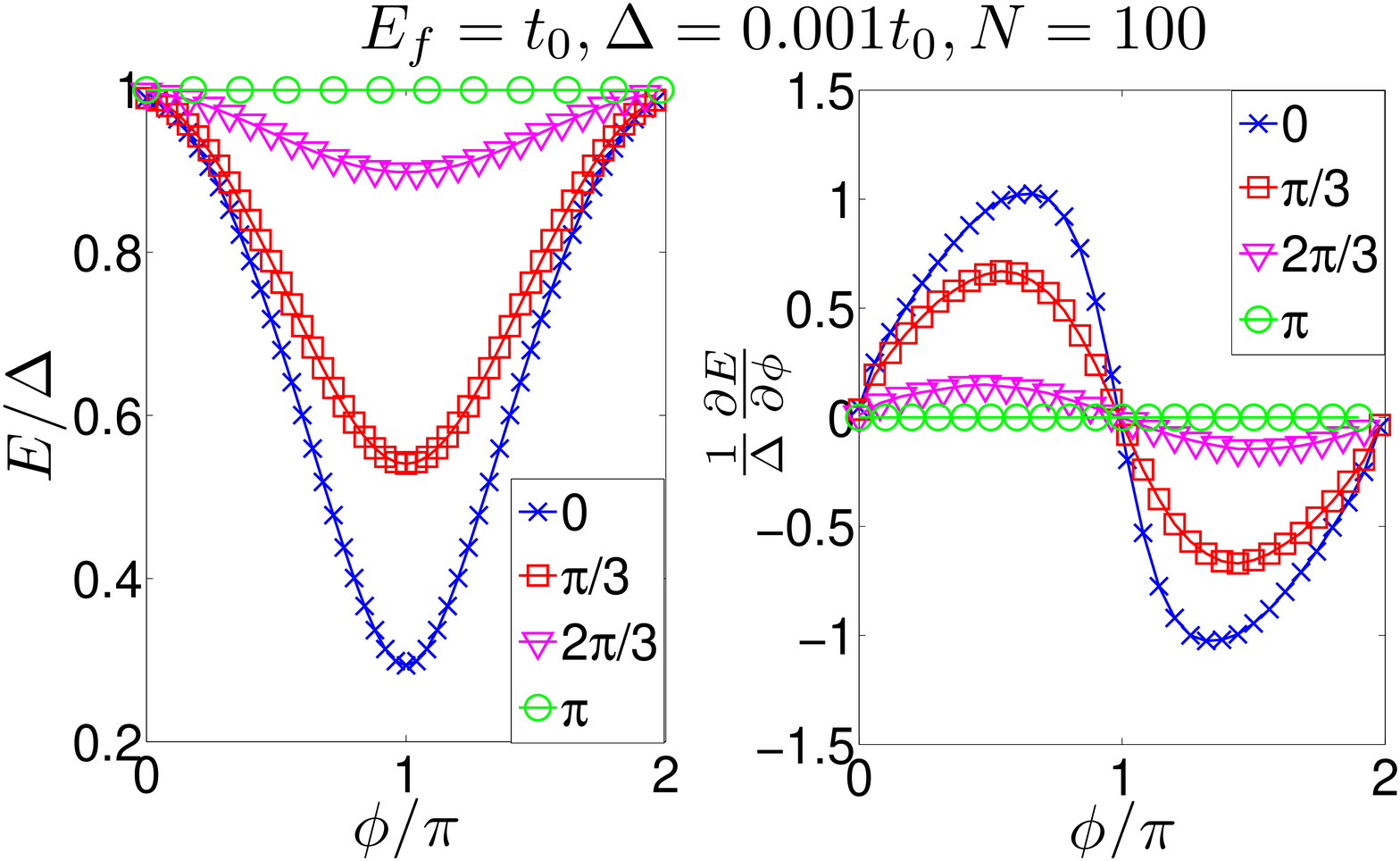}
 \hfill
\includegraphics[width=0.9\columnwidth]{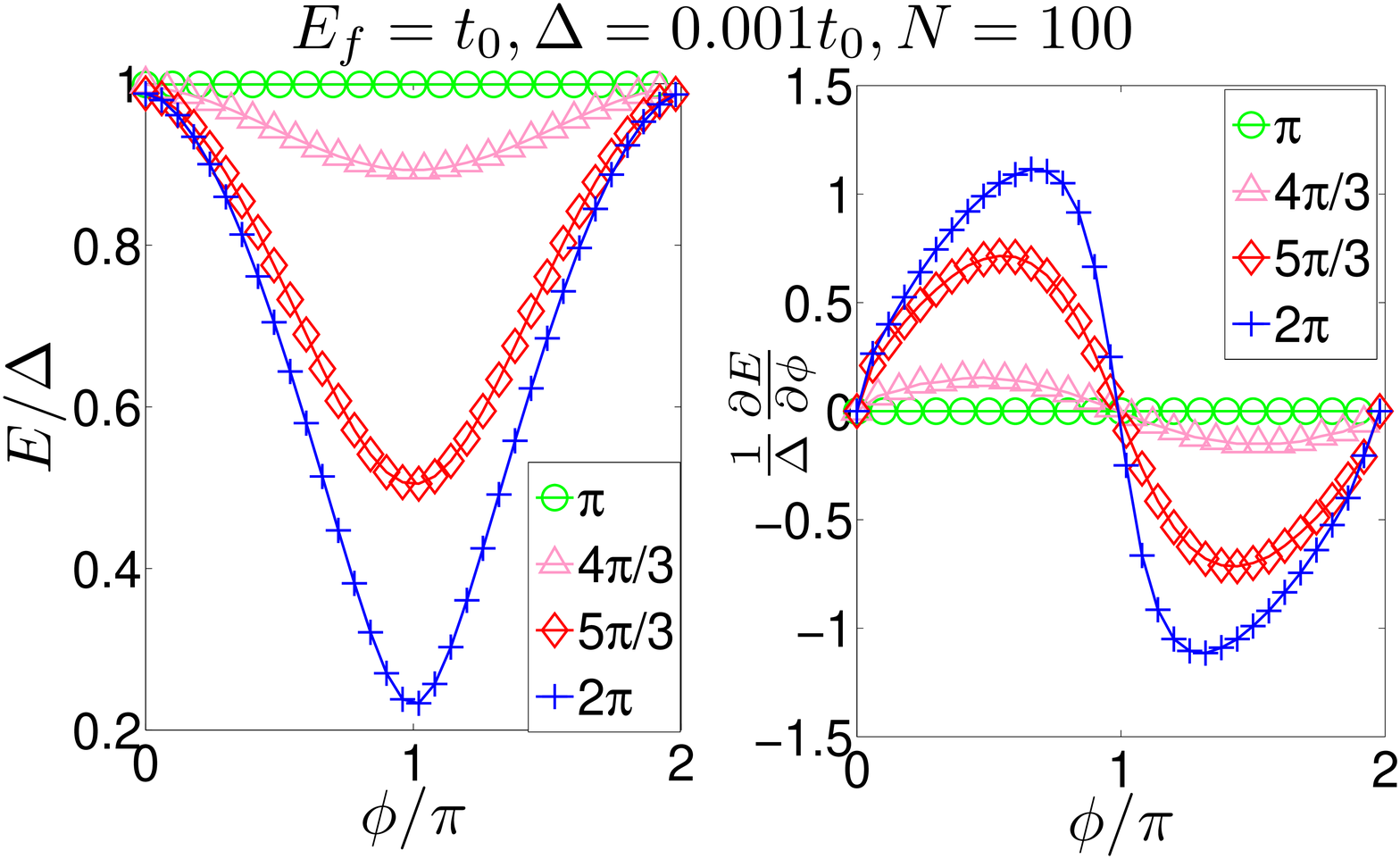}
\caption{The Andreev levels (left), $E/\Delta$, and 
normalized Josephson currents (right), $(1/\Delta\partial) E/\partial \phi$, 
vs. superconducting phase for different SOI strength measured by AC phases($w_{so}\pi$).
Here $N=100$, $E_f=t_0$, $\Delta=0.001t_0$, and $\xi_0=500\sqrt{3} a$. }
\label{fig4}
\end{figure}
\begin{figure}[ht]
\centering
\includegraphics[width=0.8\columnwidth]{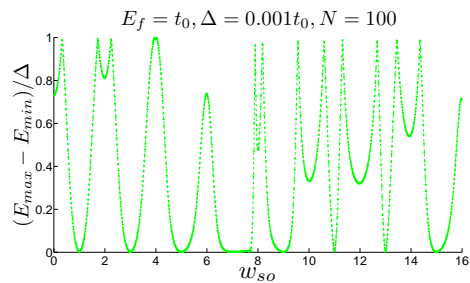}
\caption{Difference of highest and lowest Andreev levels
 vs. SOI strength measured by $w_{so}$. The difference is proportional to the critical current.
 Here $N=100$, $E_f=t_0$, $\Delta=0.001t_0$, and $\xi_0=500\sqrt{3} a$. }
 \label{fig6}
\end{figure}

The Andreev levels are presented in Fig.~\ref{Al1} as a function of the superconducting phase
difference $\phi$ for different SOI strength in the 
dimensionless variables $E/\Delta$, $w_{so}$, and $\phi/\pi$. When the AC phase is equal to $(2n+1)\pi$, the energy
of the Andreev states is $\Delta$ and independent of $\phi$. Since the Josephson current
is proportional to the first derivative of the energy $E$ of the Andreev bound state 
with respect to the phase difference $\phi$,
this implies that the zeros of the critical current occurs at the values $(2n+1)\pi$ of the AC phase. 
This conclusion is
consistent with our theoretical result in Eq.~(\ref{Andreev level}). In
Fig.~\ref{fig4}, we show the Andreev levels and the normalized Josephson currents, $(1/\Delta\partial) E/\partial \phi=-I/(2e\Delta/\hbar)$, associated
with the different AC phases vs. superconducting phase difference. The results show that the
Josephson currents  amplitudes are tuned by the AC phase. Under the short junction limit, the energy of the
Andreev levels is always equal to the superconducting gap $\Delta$
when the phase difference $\phi=0$ and minimum when $\phi=\pi$ for
any SOI strength. Therefore the value of $E(0)-E(\pi)$ is a direct indicator of the
amplitude of the critical current. $E(\pi)-E(0)$ versus AC phase, $\phi_{AC}=w_{so}\pi$, is
shown in Fig.~\ref{fig6}. The zeros  correspond to the value
$(2n+1)\pi$ of the AC phase and the peaks are not always corresponding
to $2n\pi$ but show a doubled peaked structure centered around $2n\pi$. 
This is because of weak
localization effect,\cite{Anderson} similar with the AC phase effect on
conductance oscillation in normal junctions.\cite{Nikolic,Mario}

\begin{figure}[ht]\label{long1}
\centering
\includegraphics[width=0.8\columnwidth]{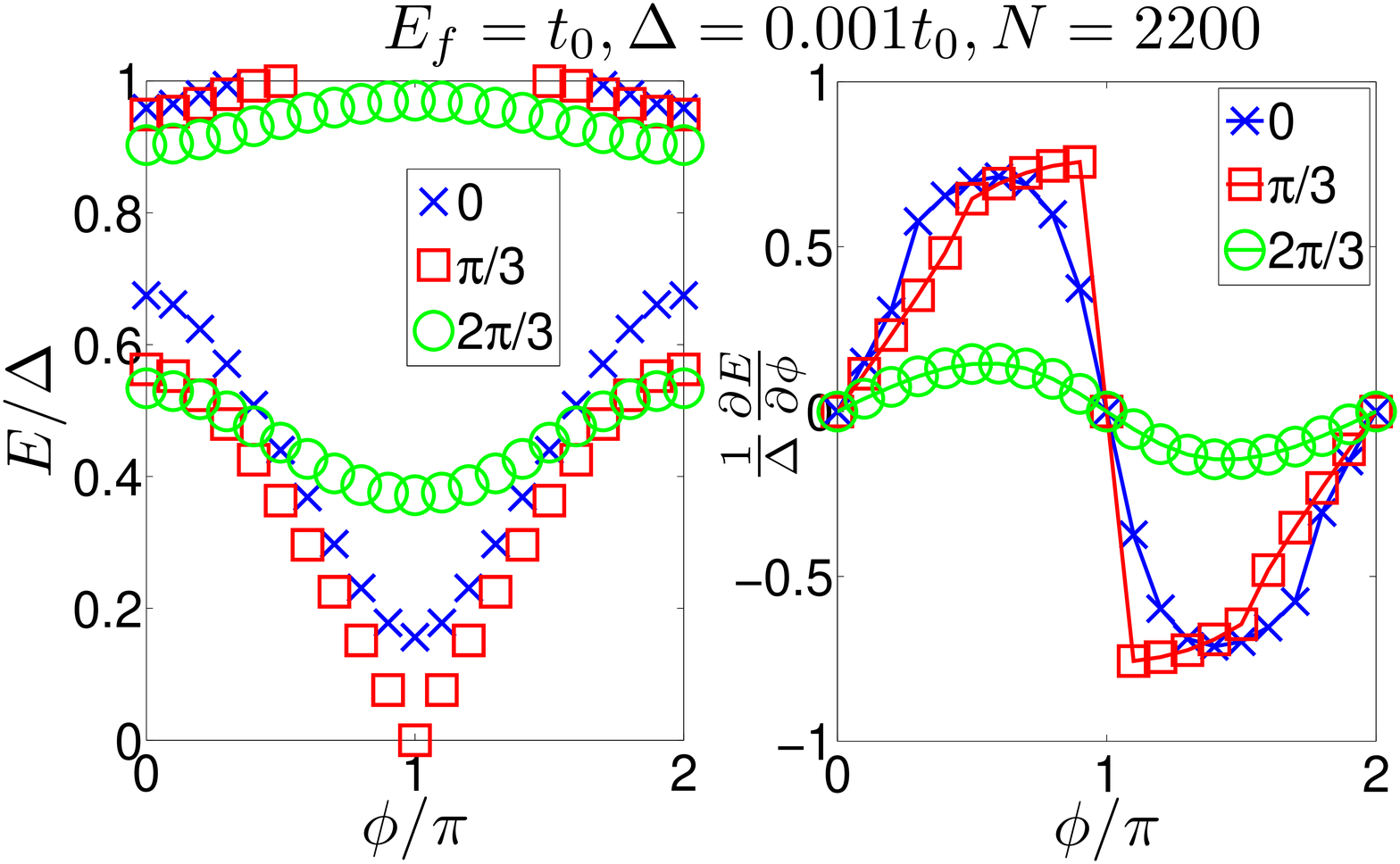}
\hfill
\includegraphics[width=0.8\columnwidth]{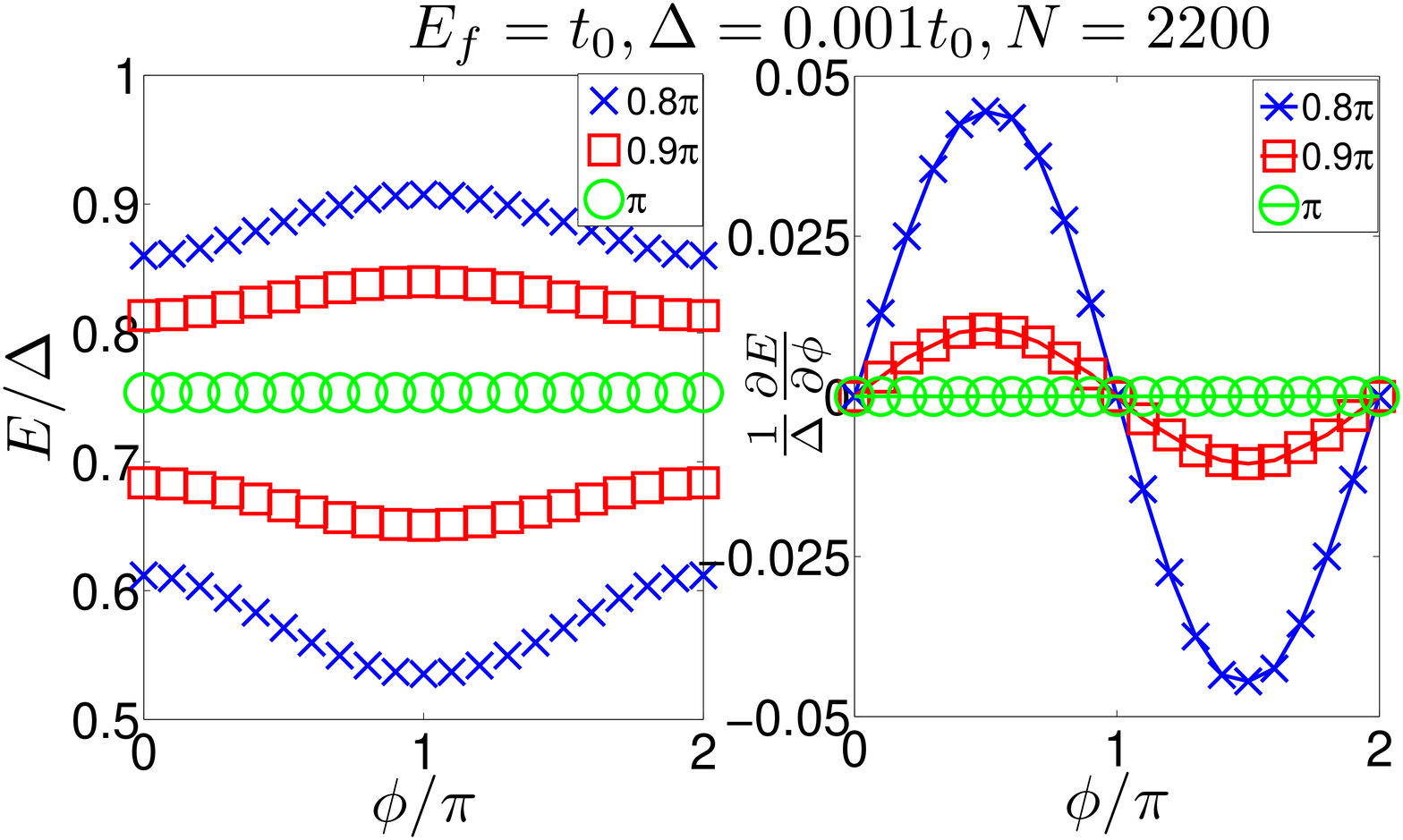}
\caption{Andreev levels as a function of superconducting phase difference
 $\phi$ for different SOI strength measured by AC phase($w_{so}\pi$). Here $N=2200$, $E_f=t_0$, $\Delta=0.001t_0$, and $\xi_0=500\sqrt{3} a$.}
 \label{fig7}
\end{figure}

We next consider the long channel length limit, $L\geq\xi_0$, where not only the
transmission function but also the phase shift due to the AC effect will
affect the Andreev levels. As an example, Fig.~\ref{fig7} plots the Andreev levels
as a function of superconducting phase difference $\phi$ with different AC phases in
a long-junction. Here the number of sites in the ring is changed from
$100$ to $2200$, {\em i.e.} the length of the ring changes by a factor of 11, while the
 other parameters are unchanged. The Andreev levels
are shifted due to SOI and the Josephson current 
vanishes when $w=(2n+1)\pi$ as in the case of the  short-junction
and confirms that the backscattering will not affect the zeros of the critical current 
in the Josephson junction in both the short and long channel length limits. 

\begin{figure}[ht]\label{short2}
\centering
\includegraphics[width=0.8\columnwidth]{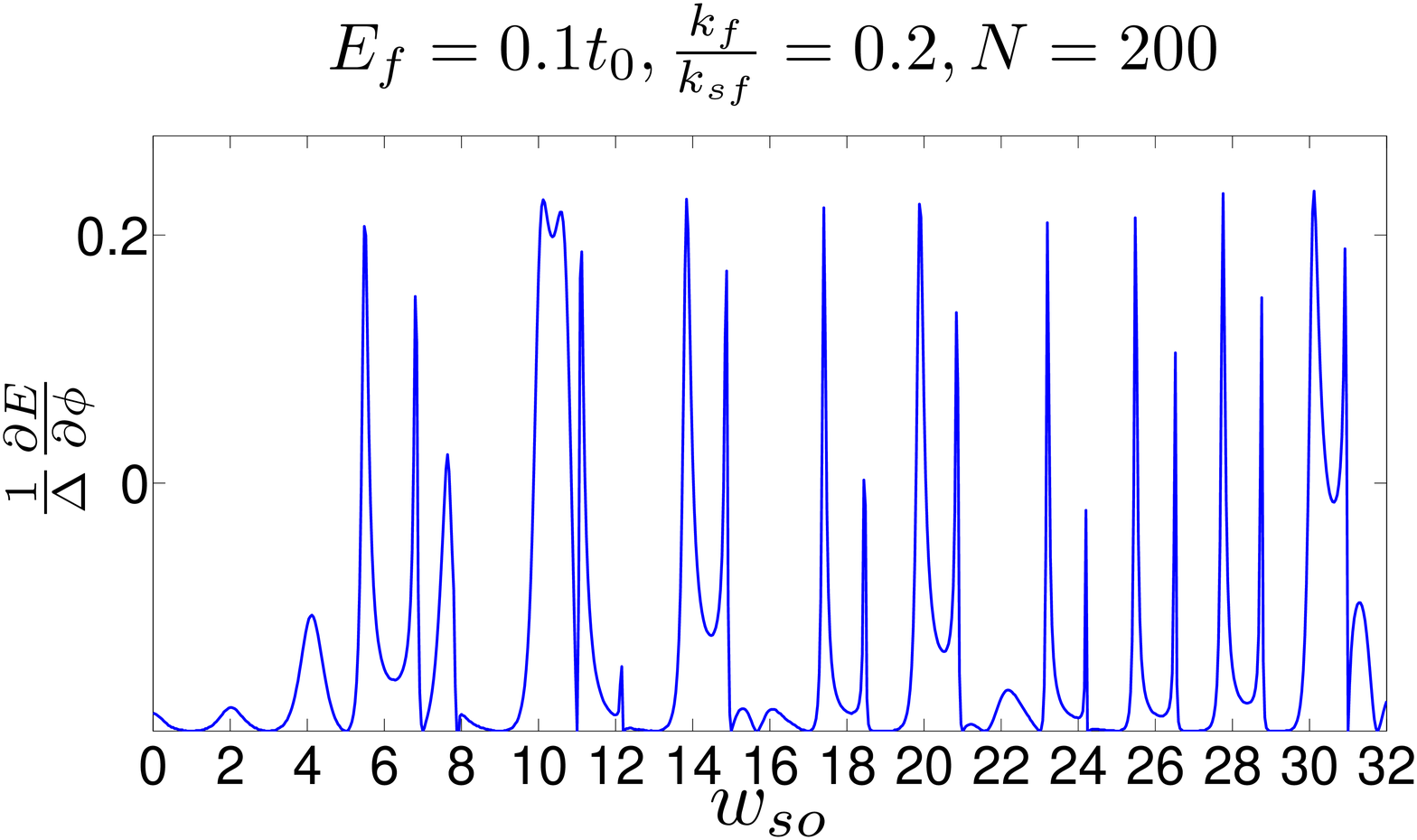}
\includegraphics[width=0.8\columnwidth]{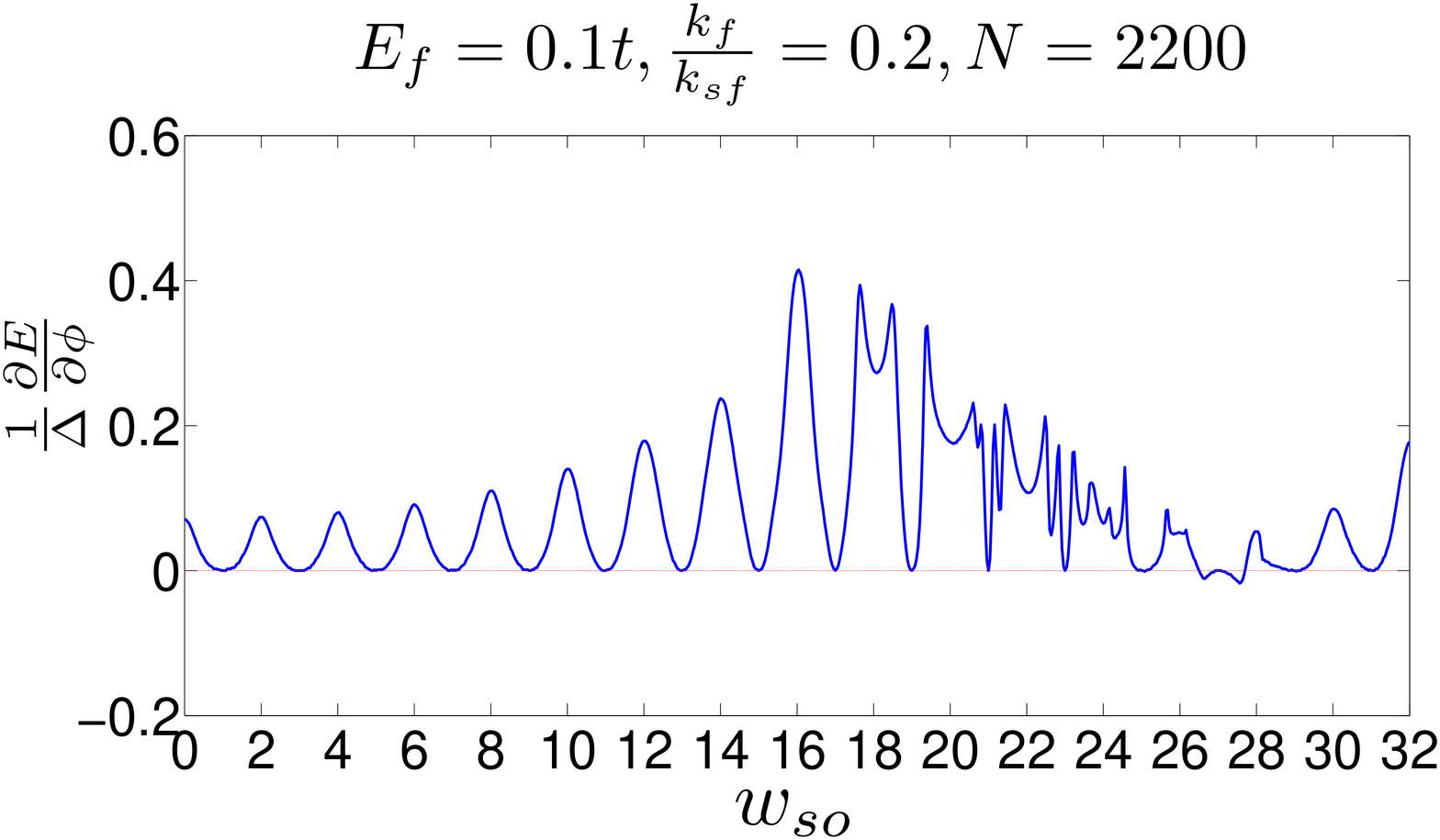}
\includegraphics[width=0.8\columnwidth]{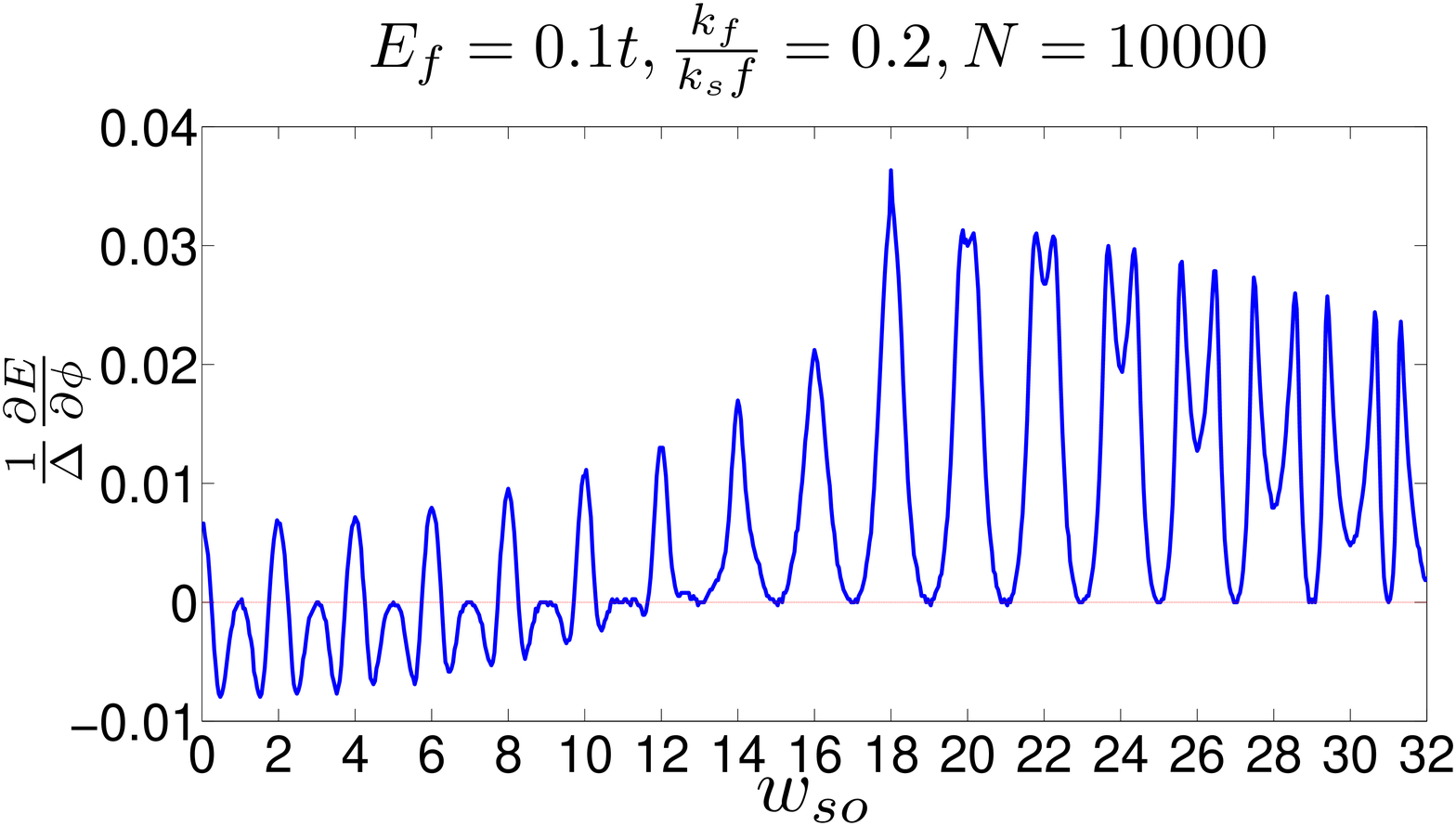}
\caption{Normalized criticla current vs. the SOI strength measured by $w_{so}$ in different lengtgh scale of the SOI ring. Here the Fermi energy $E_f=0.1t_0$, the Fermi wavelength mismatch between the SOI ring and the superconducting leads is $\frac{k_f}{k_{sf}}=0.2$ and $N=200$.}
\label{fig8}
\end{figure}

When the Schottky barrier and the Fermi wavelength mismatch are also considered, the current-SOI relation
will be more complicated because the electon transmission function will be strongly affected by these
factors. However our numerical resutls show that the zeros of the critical current are the same
to the analytical result Eq.~(\ref{JC}). 
We explore the Josephson current oscillations due to the AC phase at the lower, more realistic, Fermi
energy  $E_f=0.1t_0$, and take $\Delta=0.0002t_0$, $\xi_0=500\sqrt{5} a$. The density of the electron in the experiment using
HgTe hetero-structures
 varied from $n_{2D}=1.85\times 10^{12} {\rm cm}^{-2}$ to $n_{2D}=2.21\times 10^{12} {\rm cm}^{-2}$ and the effective mass is $m^*=0.031m_0$.\cite{AC experiment1} 
To make our parameters match the experimental data, by using the relation $E_f=0.1t_0=\hbar^2/2m^*a^2=\hbar^2 k_f^2/2m^*$, 
$\pi n_{2D}=k_f^2$, the parameter $a$ is required to be around $1.5$ nm. When we choose $N=2200$, the perimeter of the ring in our calculation is $3.3\mu$m
 which is the same order to the $r_0=1 \mu$m in the experiment.\cite{AC experiment1}
 The superconducting gap $\Delta=0.0002t_0=0.002 E_f\approx 1.72 $ K which is the same order of the gap of the conventional superconductor 
 such as Al where $T_c=1.18$ K.\cite{sup}  To capture the effects of the Schottky barrier and the Fermi wavelength mismatch on the short and
long Josephson junction more clear, we explore three ring length scales $L=200a$, $L=2200a$, and $L=10000a$, shown in Fig.~\ref{fig8}. 
The Schottky barrier is chosen to be $0.8t_0$ and the Fermi wave vector mismatch is $k_f/k_{sf}=0.2$ where $k_f$ and $k_{sf}$
are corresponding to the Fermi wavevector in the normal region and superconducting leads.
The above three oscillations have a common characteristic, when the AC phase is equal to
$(2n+1)\pi$, the critical current is zero. This means the AC effect is present even at
a large Shottky barrier and Fermi wavelength mismatch. In the bottom panel of Fig.~\ref{fig8}, the critical current has negative values, 
which would indicate that in this case there could exists a $\pi$ junction.\cite{rf3} This phenomena is due to the large difference of the phase 
shifts acquired by the electron and hole in the normal region which is related to the nonzero momentum $Q$ of the cooper pairs \cite{pith,rf3} and 
has been observed in the superconducting/feromagnetic/supconducting junctions.\cite{piex1,piex2,piex3} 
However the $\pi$ junction can not be realized in our model because it
requires the condition $L\gg \xi_0$ and the electron will be totally decohered and the Josephson current will not survive.

The advantage of observing the AC effect in the ring Josephson junction lays on
the fact that the dephasing electrons do not contribute to the
Josephson current. The dephasing effect due to the impurities can be considered
through the Green's function method. The tunneling Hamiltonian used in calculating Josephosn current is given by:\cite{tunneling}
\begin{eqnarray}\label{hamil1}
H&=&H_R+H_L+H_T \nonumber \\
H_T&=&\sum_{kq}(T_{kq\uparrow}C_{q\uparrow}^+
C_{k\uparrow}+T_{kq\downarrow}C_{q\downarrow}^+
C_{k\downarrow}+T^*_{kq\uparrow}C_{k\uparrow}^+
C_{q\uparrow}\nonumber \\
&&+T^*_{kq\downarrow}C_{k\downarrow}^+
C_{q\downarrow})
\end{eqnarray}
where $C_k(C^+_k)$ is the annihilation(creation) operator in left side,
$C_q(C_q^+)$ is the one in right side, and $T_{kq\uparrow}$ is the tunneling amplitude of the spin up electron from the left lead
with the momentum $k$ to the right lead with momentum $q$. 
In the short junction limit $L\ll\xi$, we can assume $T_{kq}$ to be a constant $T$, and the Josephson current is calculated from\cite{micro}
\begin{eqnarray}\label{GF-+}
I=\frac{8e}{h}\int_{-\infty}^{\infty}
d\omega Im\left[\frac{(
T\hat{g}_L^{r}(\omega)T^*\hat{g}_R^{r}(\omega))_{11}\}}{D^r(\omega)}\right]
\end{eqnarray}
where $D^r(\omega)={\rm Det}[1-|T|^2\tau_3\hat{g}_L^{r}(\omega)\tau_3\hat{g}_R^{r}(\omega)]$ and $\tau_3$ is the usual Pauli matrix.
This method gives the same analytical form of the Josephson current as in  Eq~(\ref{Andreev level}). However, when $L>\xi$,
the parameters of the normal material between the two superconducting leads will affect the tunneling amplitude $T_{kq}$ so strongly
that the assumption of the constant $T_{kq}$ is not valid any more. In this case we write our  tunneling Hamiltonian as:
\begin{eqnarray}\label{hamil2}
H&=&H_R+H_L+H_n+H_T \nonumber \\
H_T&=&\sum_{kq}(A_{kn}C_{k\sigma}^{\dagger}(r_L)
C_{-k-\sigma}^{\dagger}(r_L)d_{n'-\sigma}(r_L)d_{n\sigma}(r_L)+\nonumber \\
&&(A_{qn}C_{q\sigma}^{\dagger}(r_R) C_{-q-\sigma}^{\dagger}(r_R)d_{n'-\sigma}(r_R)d_{n\sigma}(r_R)+H.C\nonumber \\
H_n&=&E(n)d_n^{\dagger}d_n,
\end{eqnarray}
where $d_n^{\dagger}(d_n)$ is the creation(annihilation) operator of the electron in the normal material and the index $n$ includes
all the quantum numbers such as momentum and spin, $H_n$ is the Hamiltonian of the normal material, $E(n)$ is the eigenvalule of
$H_n$, $A_{kn}$ is the transmission amplitude on the left $S/N$ interface. $H_T$ in Eq.~(\ref{hamil2}) describes the Andreev reflection and can allow
us to consider the effect of the phase shifts acquired in the normal region. The current in this case is modified to read
\begin{eqnarray}\label{GF-+2}
I=\frac{8e}{h}\int_{-\infty}^{\infty}
d\omega Im\left[\frac{(T_n\hat{g}_L^{r}(\omega)T^*_{n'}\hat{g}_R^{r}(\omega))_{11}\}}{D^r(\omega)}\right],\nonumber \\
T_n=A_{kn}\hat{G}^{r}(r_R,r_L,n),T^*_{n'}=A^*_{kn}\hat{G}^{a}(r_R,r_L,n'),
\end{eqnarray}
where $\hat{G}^{r(a)}(r_R,r_L,n,\sigma)$ is the retarded (advanced) Green's function of the electron with energy $E(n)$ and spin $\sigma$ in the normal
region. When $r_R-r_L=L\ll\xi_0$, $\hat{G}^{r(a)}(r_R,r_L,n)$ is close to the Fermi distribution $f(E(n))$ and Eq.~(\ref{GF-+2}) has the same form to its
short junction limit Eq.~(\ref{GF-+}). However when $r_R-r_L=L\sim\xi_0$, because $E(n)=-E(n')$, $\hat{G}^{r}(r_R,r_L,n)$ and $\hat{G}^{a}(r_R,r_L,n')$ are
not conjugate to each other. As a result, even the product $\hat{G}^{r}(r_R,r_L,n)\hat{G}^{a}(r_R,r_L,n')$ from the trajectories with an
identical ensemble of scattering centers, which is corresponding to the ladder correction, is complex with a random phase instead of real.
Therefore the ladder correction due to the impurity scattering is zero in the Josephson junction current. This is very different to the calculation
of the conductance which is proportional to $|\hat{G}^{r(a)}(E=E_f)|^2$ where the ladder correction due to the impurities is nonzero. This
is why impurity scattering will not affect the Josephson current. This can be confirmed from the experimental data of
the Fraunhofer-like interference pattern of the critical current in the Josephson junctions.\cite{1982book}
In those experiments, an oscillation with an amplitude of almost
$100\%$ of the total current were observed.\cite{1982book} Similarly 
in the ring Josephson junction with SOI, an oscillation due to the AC phase with a larger
amplitude than that in the conductance experiments is expected.\cite{AC experiment1} Especially if a
one channel ring limit is achieved, a $100\%$ oscillation due to the AC
phase should be observed. When the normal material is
semiconducting, a tunable Josephson current controlled by a voltage
can be obtained creating a Josephson field effect
transistor,\cite{JoFET1,JoFET2,JoFET3} where the ring Josephson junction can
be switched on and off by tuning the AC-phase and carrier concentration. 
Recently, a voltage of $-70$ V has been
reported to switch off a Al/InAs/Al Josephson junction.\cite{JoFET3} Our result
provides a new possible way to control the Josephson current by a
gate voltage which can be as small as several meV \cite{AC
experiment1} and much smaller than the gate voltage in normal Josephson
field effect transistors.\cite{JoFET1,JoFET2,JoFET3}

\section{Experiment proposal}

\begin{figure}[ht] \centering
\includegraphics[width=0.8\columnwidth]{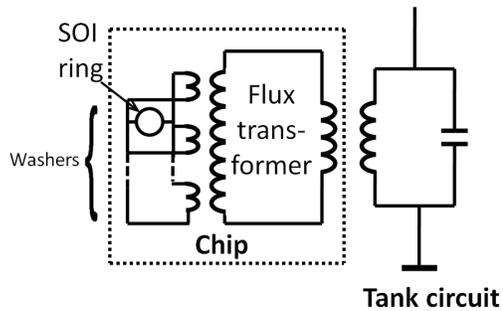}
\caption{Proposed experimental set-up.}
\label{set}
\end{figure}

In this section, we discuss the feasibility of experiments to observe AC phase through the oscillation of the Josephson current.
Our experimental proposal is based on the radio-frequency (rf) method\cite{rf1,rf2,rf3} which is a reliable method to measure
the current-phase relation (C$\Phi$R) of the Josephson current\cite{rf4,rf5}, especially for the small critical current even less
than $50$ nA.\cite{rf4}. Therefore this technique makes it possible to detect C$\Phi$R in a S/Sm/S junction although there is a
large Schottky barrier on the S/N interface. Our S/Sm/S junction, depicted in Fig.~\ref{ring}, is incorporated into a SQUID washer
of very small and well defined inductance $L$, which is coupled inductively to a high quality tank circuit in resonance with
quality $Q$ and coupled strength $k$ Fig.~\ref{set}.
The Josephson current $I_s$ in the ring and the phase difference $\alpha$ of the DC current and voltage in the tank circuit have the relation \cite{rf4}
\begin{eqnarray}\label{cur-exp}
&&\tan(\alpha)=\frac{k^2Q\beta f'(\phi)}{1+\beta f'(\phi)},\nonumber \\
&&I_s(\phi)=\frac{I_c}{k^2Q\beta}\int_0^{\phi} \tan \alpha(\phi) d\phi,
\end{eqnarray}
where $\beta=2\pi L I_c/\Phi_0$ is the normalized critical current, $f(\phi)=I_s(\phi)/I_c$ is the normalized current-phase relation and $I_c$
is the critical current. Therefore, we can observe the oscillation of the Josephson current induced by AC phase by detecting the phase
difference of the current and voltage in the tank circuit. According to the amplitude of the Josephson currents and the size of the Josephson
junctions in Refs. \onlinecite{rf5,exp1,exp2} we can expect a Josephson current up to $I_c=1$ $\mu $A in the ring shape structure whose radius is $1$
$\mu $m and width of each arm is $300$ nm. If we choose the inductance of the superconducting loop $L=80p$H,\cite{rf3} $k^2Q=0.1$, the
normalized critical current $\beta=2\pi L I_c/\Phi_0=0.25$, the phase difference $\alpha$ of the current and voltage in the tank circuit will
be in the range $(-0.02 {\rm rad},0.02 {\rm rad})$.

\section{Conclusions}

In this work we have studied the interplay between the critical Josephson current in a ring  junction and the AC effect. 
We have calculated the Andreev levels in
a Rashba type SOI ring system attached to two superconducting leads both analytically and numerically.
Numerically, using the boundary condition of the multiple terminal junction in the
tight-binding model, we calculate the Andreev levels and the Josephson currents
in both short and long channel length limit. After considering the backscattering in the SOI ring, large Schottky barrier on the S/N interface,
Fermi wavelength mismatch between the superconducting leads and semiconducting ring, and the effect of the impurities in the ring, 
we find that the oscillations of the Josephson current due to the AC phase are robust. The amplitude oscillation of the Josephson current
due to the AC effect in the D-C Josephson junction is expected to be larger than the conductance oscialltions observed in 
normal ring structures without superconducting elements. 
These results suggest an alternative and likely better way to observe the AC phase.
Also,  since one period oscillation of the AC phase only needs several meV, 
we provide a possible way to create Josephson junction FETs controlled  by a voltage of  the order of meV.

\section*{Acknowledgment}
This work was supported support from
ONR under Grant No. ONR-N000140610122, by NSF
under Grant No. DMR-0547875, by the Research Corporation Cottrell Scholar Award, and by SWAN-NRI.

\appendix

\section{The quasiparticle state within the gap}\label{sing}
We presents the detail discussion of the quasiparticle states in the gap
in this appendix. As we know that the quasiparticle states in the
superconductor can be described by the Bogoliubov equation
\begin{eqnarray}\label{bogo}
\left(\begin{array}{cc}H(r)&\Delta(r)\\\Delta^*&-H(r)\end{array}\right)\left(\begin{array}{c}u_k(r)\\v_k(r)\end{array}\right)=E\left(\begin{array}{c}u_k(r)\\v_k(r)\end{array}\right),
\end{eqnarray}
where $H(r)$ is the one-electron Hamiltonian, defined as
\begin{eqnarray}\label{he}
H(r)=-\frac{\hbar^2}{2m^*}\nabla^2+U(r)-\mu,
\end{eqnarray}
where $\mu$ is the electrochemical potential, $U(r)$ is the scalar
potential and $m^*$ is the effective mass. For a homogeneous
superconductor with $\Delta(r)$=$\Delta_0$ and $U(r)=0$, the
solution of Eq.\ref{bogo} can be written as
\begin{eqnarray}\label{uv}
&&\left(\begin{array}{c}u_k(r)\\v_k(r)\end{array}\right)=\exp(ikr)\left(\begin{array}{c}u_0\\v_0\end{array}\right),
u_0=\cos(\frac{\theta}{2}), \nonumber
\\&&v_0=\sin(\frac{\theta}{2})e^{-i\phi}, E=\sqrt{(\frac{\hbar^2k^2}{2m^*}-\mu)^2+\Delta_0^*\Delta_0}
\end{eqnarray}
where $\cos\theta=(\frac{\hbar^2k^2}{2m^*}-\mu)/E$, $\phi$ is the
phase of the superconductor. $|k|>k_f$ is corresponding to
electron-like quasiparticle because in this case, $\cos\theta>0,
\theta/2<\pi/2, \cos(\theta/2)>\sin(\theta/2)$. For the similar
reason, $|k<k_f|$ is corresponding to the hole-like quasiparticle. Here
only the positive energy $E$ is considered. In 1-d case, given energy $E>\Delta$,
there are four quasiparticles, two of them are the right moving
particles and the other two are the left moving particles. When
considering $E<\Delta$, according to Eq.\ref{uv}, the wave vector
$k$ satisfies
\begin{eqnarray}\label{wsk}
\frac{\hbar^2k^2}{2m^*}-\mu=\pm i\epsilon,
\end{eqnarray}
where $\epsilon=\sqrt{\Delta^2-E^2}>0$. Therefor the wave vector $k$
has to take the form
\begin{eqnarray}\label{wsk1}
k=k_r+ik_i,
\end{eqnarray}
where the real part and imaginary part of $k$ satisfy
\begin{eqnarray}\label{wsk2}
&&\frac{\hbar^2k_r k_i}{m}=\pm\epsilon,\nonumber \\
&&\frac{\hbar^2}{2m^*}(k_r^2-k_i^2)-\mu=0.
\end{eqnarray}
 Given energy
$E<\Delta$, the wave vectors of the quasiparticles in this case are
corresponding to the four solutions of Eq.\ref{wsk2}. The solution
of Eq.(\ref{bogo}) will be
\begin{eqnarray}\label{uv}
u_0=\frac{\sqrt{2}}{2}, v_0=\frac{E\pm
i\epsilon}{\Delta}e^{-i\phi}u_0.
\end{eqnarray}
As a result, $|u_0|=|v_0|=\frac{\sqrt{2}}{2}$.

If a normal conductor is coupled to a superconductor, a unique
reflection with energy less than gap, namely Andreev reflection, can
be observed. To calculate the scattering matrix in the interface, we
must know the input and output particles, which are corresponding to the
right moving and left moving particles separately, in the
superconducting lead. It is easy to find them when the energy is larger
than the superconducting gap. However, when the energy is less than the superconducting gap, the wave vector has
to be analytic continuous to complex plane Eq.\ref{wsk1}. It is hard
to say which solution is corresponding the left moving or the right moving particle. We will
figure out this problem by some concepts from retarded Green's
function.

Retarded Green's function in the position and energy space, say
$G(x,x',E_k)$, is viewed as two traveling wave function outward from a
source term $\delta(x-x')$. The opposite case is so called Advanced
Green's function. One way to calculate S matrix is to use retarded
Green function\cite{Data} but not Advanced Green's function because
we are interested in the case that if there is an excitation at some
point, how the excited particle travels away from the point of the
excitation. This kind of retarded Green's function of a fermion is
$\int_{-\infty}^{\infty}\frac{e^{ikx}}{E-E_k+i\delta}dk$ where
$\delta>0$, therefore we have a pole where $E_k=E+i\delta$.
Physically we hope that when $x>0$, the integral gives us the right
moving particle and when $x<0$, the integral gives us the left moving
particle.

Now let us come back to our question. In our discussion, the energy
$E$ is chosen to be positive. As a result, $E_k$ must be in the
first quadrant of a energy complex plane and just a little bit above
the real axis. Actually, Eq.(\ref{wsk1}) is a general form of wave
vector. $k_i$ is infinitely small when energy $E>\Delta$ and finite
when energy $E<\Delta$. Considering $E>\Delta$ and according to
Eq.(\ref{wsk1}), the energy of the quasiparticles can be written as
\begin{eqnarray}\label{rmee}
E_k&=&\sqrt{(\frac{\hbar^2k^2}{2m^*}-\mu)^2+\Delta_0^*\Delta_0+2i(\frac{\hbar^2k^2}{2m^*}-\mu)\frac{\hbar^2
k_r k_i}{m}}\nonumber
\\&\simeq& \sqrt{(\frac{\hbar^2k^2}{2m^*}-\mu)^2+\Delta_0^*\Delta_0}+i(\frac{\hbar^2k^2}{2m^*}-\mu)\frac{\hbar^2
k_r k_i}{m},
\end{eqnarray}
here $k_i$ is infinitely small and we neglect the $k_i^2$ term. Since
$E_k$ is in the first quadrant of the complex plane,
$(\frac{\hbar^2k^2}{2m^*}-\mu)\frac{\hbar^2 k_r k_i}{m}>0$. For the
right moving electron-like quasiparticle $\Psi_{re}$, since
$(\frac{\hbar^2k^2}{2m^*}-\mu)>0$ and $k_r>0$, $k_i$ must be also
larger than zero, say the wave vector of the right moving electron-like
quasiparticle of the first quadrant in complex wave vector plane.
For the similar reason we have the results that the left moving
electron-like quasiparticle $\Psi_{le}$ is in the third quadrant, the right
moving hole-like quasiparticle $\Psi_{rh}$ is in the second quadrant
and the left moving hole-like quasiparticle $\Psi_{lh}$ is in the fourth
quadrant.

Although we get this result in the case $E>0$, it is also valid when
$E<0$ because for the wave vectors, the case of $E<\Delta$ is just
the analytical continuous of that of $E>\Delta$. The wave functions
of these four quasiparticles within the superconducting gap are
\begin{eqnarray}\label{wfig}
&&\Psi_{re}=\left(\begin{array}{c}\frac{\sqrt{2}}{2}\\\frac{\sqrt{2}}{2}e^{-i\alpha}e^{-i\phi}\end{array}\right)e^{i(|k_r|+i|k_i|)x},\nonumber\\
&&\Psi_{le}=\left(\begin{array}{c}\frac{\sqrt{2}}{2}\\\frac{\sqrt{2}}{2}e^{-i\alpha}e^{-i\phi}\end{array}\right)e^{-i(|k_r|+i|k_i|)x},\nonumber\\
&&\Psi_{rh}=\left(\begin{array}{c}\frac{\sqrt{2}}{2}\\\frac{\sqrt{2}}{2}e^{i\alpha}e^{-i\phi}\end{array}\right)e^{i(-|k_r|+i|k_i|)x},\nonumber\\
&&\Psi_{lh}=\left(\begin{array}{c}\frac{\sqrt{2}}{2}\\\frac{\sqrt{2}}{2}e^{i\alpha}e^{-i\phi}\end{array}\right)e^{i(|k_r|-i|k_i|)x},
\end{eqnarray}
where $\alpha=\arccos(E/\Delta)$.

An interesting conclusion should be noticed that in both the left and
right superconducting leads, exponentially decay quasiparticles are
always the input particles and exponentially increase quasiparticles are
always output particles. The velocity
of the electrons and the current carried by these electrons can be calculated
through the velocity operator and electron current operator
\begin{widetext}
\begin{eqnarray}\label{vi}
v=<\Psi|\hat{v}|\Psi>&=&\left(\begin{array}{cc}u_0^*&
v_0^*\end{array}\right)e^{-i(k_r-ik_i)x}\left(\begin{array}{cc}-i\hbar\frac{\partial}{\partial
x}&0\\0&i\hbar\frac{\partial}{\partial
x}\end{array}\right)e^{i(k_r+ik_i)x}\left(\begin{array}{c}u_0\\v_0\end{array}\right)=0,\\
I=<\Psi|\hat{I}|\Psi>&=&e\left(\begin{array}{cc}u_0^*&
v_0^*\end{array}\right)e^{-i(k_r-ik_i)x}\left(\begin{array}{cc}-i\hbar\frac{\partial}{\partial
x}&0\\0&-i\hbar\frac{\partial}{\partial
x}\end{array}\right)e^{i(k_r+ik_i)x}\left(\begin{array}{c}u_0\\v_0\end{array}\right)\nonumber\\
&=&e\frac{\hbar k_r}{m^*}e^{-2k_ix},
\end{eqnarray}
where $e$ is the charge of an electron. The velocity of these decay
quasiparticles are zero but the current carried by these are not
zero. Although Eq.(\ref{vi}) shows that the current is decay, since the
quasiparticle will decay to cooper pair\cite{BTK} which can carry a
supercurrent, the total current due to one quasiparticle is
$I=e\frac{\hbar k_r}{m^*}$

\section{spin-orbit coupling in a tight binding model ring system}
There have been many theoretical papers talking about the Rashba
interaction in a ring system, such as the eigenenergy, wavefunction and
so on by the analytical method in the continuous case and the exact
transmission function through the numerical calculation in the tight binding
model. However the eigenenergy and wavefunction in the tight binding model is
still undiscussed. Although this is easy to be derived, we write down
the conclusion briefly to make our paper more readable.

The Hamiltonian of the SOI ring system has been given in Eq\ref{tbra}. Now
we give this tight binding model Hamiltonian of the electron around the
point $\varphi=0$.
\begin{eqnarray}\label{tighth}
H=\left(\begin{array}{cccccc}2t_0&0&-t_0&it_{so}e^{i\delta\varphi/2}&0&0\\0&2t_0&it_{so}e^{-i\delta\varphi/2}&-t_0&0&0\\
-t_0&-it_{so}e^{i\delta\varphi/2}&2t_0&0&-t_0&it_{so}e^{-i\delta\varphi/2}\\
-it_{so}e^{-i\delta\varphi/2}&-t_0&0&2t_0&it_{so}e^{i\delta\varphi/2}&-t_0\\
0&0&-t_0&-it_{so}e^{-i\delta\varphi/2}&2t_0&0\\
0&0&-it_{so}e^{-i\delta\varphi/2}&-t_0&0&2t_0\end{array}\right)
\end{eqnarray}
\begin{eqnarray}\label{eigenf}
\Psi_{e,n}=\left(\begin{array}{c}\sin(\frac{\gamma}{2})e^{-i(n-1)\delta\varphi}\\
\cos(\frac{\gamma}{2})e^{-in\delta\varphi}\\\sin(\frac{\gamma}{2})\\\cos(\frac{\gamma}{2})\\\sin(\frac{\gamma}{2})e^{i(n-1)\delta\varphi}\\
\cos(\frac{\gamma}{2})e^{in\delta\varphi}\end{array}\right),
\Psi_{e,m}=\left(\begin{array}{c}\cos(\frac{\gamma}{2})e^{-in\delta\varphi}\\
-\sin(\frac{\gamma}{2})e^{-i(n+1)\delta\varphi}\\\cos(\frac{\gamma}{2})\\-\sin(\frac{\gamma}{2})\\\cos(\frac{\gamma}{2})e^{in\delta\varphi}\\
-\sin(\frac{\gamma}{2})e^{i(n+1)\delta\varphi}\end{array}\right).
\end{eqnarray}

The eigenfunction of this tight binding model Hamiltonian is the
same as the eigenfunction of the continuous Hamiltonian. Acting the tight
binding model Hamiltonian Eq.\ref{tighth} on the wave function
Eq.\ref{eigenf} and focusing on $\varphi=0$ gives us
\begin{eqnarray}\label{thw}
\hat{H}\Psi_{e,n}&=&\left(\begin{array}{cc} 2t_0(1-\cos(n-1)\delta\varphi)&2t_{so}\sin(n-\frac{1}{2})\delta\varphi\\
2t_{so}\sin(n-\frac{1}{2})\delta\varphi&2t_0(1-\cos(n\delta\varphi))
\end{array}\right)\left(\begin{array}{c}\sin(\gamma/2)\\\cos(\gamma/2)\end{array}\right)\nonumber\\
&=&2t_0(1-\cos(n-\frac{1}{2})\delta\varphi)\cos(\frac{1}{2}\delta\varphi)\left(\begin{array}{c}\sin(\gamma/2)\\\cos(\gamma/2)\end{array}\right)+\nonumber
\\&&\left(\begin{array}{cc}-2t_0\sin(n-\frac{1}{2})
\delta\varphi\sin(\frac{1}{2}\delta\varphi)&2t_{so}\sin(n-\frac{1}{2})\delta\varphi\\2t_{so}\sin(n-\frac{1}{2})\delta\varphi&
2t_0\sin(n-\frac{1}{2})\delta\varphi\sin(\frac{1}{2}\delta\varphi)\end{array}\right)\left(\begin{array}{c}\sin(\gamma/2)\\\cos(\gamma/2)\end{array}\right)\\
\hat{H}\Psi_{e,m}&=&\left(\begin{array}{cc} 2t_0(1-\cos(m\delta\varphi))&2t_{so}\sin(m+\frac{1}{2})\delta\varphi\\
2t_{so}\sin(m+\frac{1}{2})\delta\varphi&2t_0(1-\cos(m+1)\delta\varphi)
\end{array}\right)\left(\begin{array}{c}\cos(\gamma/2)\\-\sin(\gamma/2)\end{array}\right)\nonumber
\\&=&2t_0(1-\cos(m+\frac{1}{2})\delta\varphi)\cos(\frac{1}{2}\delta\varphi)\left(\begin{array}{c}\cos(\gamma/2)\\-\sin(\gamma/2)\end{array}\right)+\nonumber
\\&&\left(\begin{array}{cc}-2t_0\sin(m+\frac{1}{2})
\delta\varphi\sin(\frac{1}{2}\delta\varphi)&2t_{so}\sin(m+\frac{1}{2})\delta\varphi\\2t_{so}\sin(m+\frac{1}{2})\delta\varphi&
2t_0\sin(m+\frac{1}{2})\delta\varphi\sin(\frac{1}{2}\delta\varphi)\end{array}\right)\left(\begin{array}{c}\cos(\gamma/2)\nonumber
\\-\sin(\gamma/2)\end{array}\right)\\.
\end{eqnarray}
When $\gamma=\arctan(\frac{t_{so}}{t_0\sin(\delta\varphi/2)})$, we
have
\begin{eqnarray}\label{energytb}
&&E_{e,n}=2t_0(1-\sqrt{1+(t_{so}/t_0)^2}\cos[(n-1/2)\delta\varphi+\beta])\nonumber\\
&&n_{\pm}=\left(\pm\arccos\left((1-\frac{E}{2t_0})/\sqrt{1+(\frac{t_{so}}{t_0})^2}\right)-\beta\right)/\delta\varphi+\frac{1}{2}\\
&&E_{e,m}=2t_0(1-\sqrt{1+(t_{so}/t_0)^2}\cos[(m+1/2)\delta\varphi-\beta])\nonumber\\
&&m_{\pm}=\left(\pm\arccos\left((1-\frac{E}{2t_0})/\sqrt{1+(\frac{t_{so}}{t_0})^2}\right)+\beta\right)/\delta\varphi-\frac{1}{2}
\end{eqnarray}
where
$\beta=\arccos(\frac{cos(\delta\varphi/2)}{\sqrt{1+(t_{so}/t_0)^2}})$,
$n_+$ and $m_+$ are corresponding to the counterclockwise rotation
electron and $n_-$ and $m_-$ are corresponding to the clockwise rotation
electron.
\end{widetext}

We next show the equivalent between the continous limit and the tight bind model by giving the detail proof of our statement that the term $w_{so}=(2\beta/\delta\varphi-1)$ is equal to $(\sqrt{1+Q_R^2}-1)$ 
in the limit $a \rightarrow 0$. According to the L' Hopital's rule, the term $(2\beta/\delta\varphi-1)$ satisfies
\begin{eqnarray}\label{L'hopital}
\lim_{a\rightarrow 0} \frac{2\beta}{\delta\varphi}&=&\frac{f(a)}{g(a)}=\lim_{a\rightarrow 0}\frac{f'(a)}{g'(a)},\nonumber \\
f(a)&=&2\arccos(\frac{cos(\delta\varphi/2)}{\sqrt{1+(t_{so}/t_0)^2}}),\nonumber \\
&=&2\arccos(\frac{cos(a/2r)}{\sqrt{1+(\alpha m a/\hbar^2)^2}}),\nonumber \\
g(a)&=&\delta\varphi=a/r.
\end{eqnarray}
because $\lim_{a\rightarrow 0} \beta=0$ and $\lim_{a\rightarrow 0} \delta \varphi=0$. If we define $u(a)=\frac{cos(a/2r)}{\sqrt{1+(\alpha m a/\hbar^2)^2}}$,
we have \begin{eqnarray}
\label{f'a1}
f'(a)&=&\frac{d \arccos(u)}{d a}=-\frac{2}{\sqrt{1-u^2}}\frac{du}{da},
\end{eqnarray}
where \begin{eqnarray}
\label{f'a2}
1-u^2&=&1-\frac{\cos^2(a/2r)}{1+(\alpha^2 m^2 a^2/\hbar^4)}\nonumber \\
&=&\frac{1+\alpha^2 m^2 a^2/\hbar^4-\cos^2(a/2r)}{1+\alpha^2 m^2 a^2/\hbar^4}\nonumber \\
&\simeq&1+\frac{\alpha^2 m^2 a^2}{\hbar^4}-(1-\frac{a}{4r^2})\nonumber\\
&=&a^2(\frac{m^2\alpha^2}{\hbar^4}+\frac{1}{4r^4}), 
\end{eqnarray}
\begin{eqnarray}
\label{f'a3}
\frac{du}{da}&=&-\frac{1/2r\sin(a/2r)}{\sqrt{1+\alpha^2 m^2 a^2/\hbar^4}}-\frac{\cos(a/2r)\alpha^2 m^2 a/\hbar^4}{(1+\alpha^2 m^2 a^2/\hbar^4)^{3/2}}\nonumber \\
&\simeq&-a(\frac{1}{4r^2}+\frac{\alpha^2 m^2}{\hbar^4}).
\end{eqnarray}
By inserting Eq.(\ref{f'a2},\ref{f'a3}) to Eq.(\ref{f'a1}), we find that
\begin{eqnarray}\label{f'a}
\lim_{a\rightarrow 0} f'(a)&=&\frac{2}{a\sqrt{\alpha^2m^2/\hbar^4+1/4r^2}}a(\frac{\alpha^2 m^2}{\hbar^4}+\frac{1}{4r^2})\nonumber \\
&=&2\sqrt{\frac{1}{4r^2}+\frac{\alpha^2 m^2}{\hbar^4}}.
\end{eqnarray}
Substituting Eq. (\ref{f'a}) to Eq. (\ref{L'hopital}), we obtain the form
\begin{eqnarray}\label{w-Q}
\lim_{a\rightarrow 0}\frac{2\beta}{\delta \varphi}&=&\frac{f'(a=0)}{g'(a=0)}=\frac{2\sqrt{1/4r^2}+\alpha^2 m^2 a/\hbar^4}{1/r}\nonumber\\
&=&\sqrt{1+\frac{4\alpha^2 m^2 r^2}{\hbar^4}}=\sqrt{1+Q_R^2},
\end{eqnarray}
which is exactly the same as our statement.


\noindent\\  \\


\begin{thebibliography}{99}

\bibitem{A-C} Y. Aharonov and A. Casher, Phys. Rev. Lett. 53, 319 (1984).

\bibitem{AC experiment1} M. K\"{o}nig, A. Tschetschetkin, E. M. Hankiewicz, Jairo Sinova, V. Hock, V. Daumer, M. Sch\"{a}fer, C. R. Becker,
H. Buhmann, and L.W. Molenkamp, Phys. Rev. Letters, 96,076804(2006).

\bibitem{AC experiment2}  B. Habib, E. Tutuc, and M. Shayagen, Appl. Phys. Lett. 90, 152104 (2007)

\bibitem{AC} Diego Frustaglia and Klaus Richter, Rhys. Rev. B 69,
              235310(2004).

\bibitem{Nikolic} Satofumi Souma and Branislav K. Nikoli\'{c}, Phys.
                  Rev. B, 70,195346(2004).

\bibitem{path} P. Lucignano, D. Giuliano, and A. Tagliacozzo, Phys. Rev. B 76, 045324(2007).

\bibitem{AC vor1} B. J. van Wees, Phys. Rev. Lett. 65, 255 (1990).

\bibitem{AC vor2} E. Simanek, Phys. Rev. B 55, 2772 (1997).

\bibitem{AC vor3} Jonathan R. Friedman and D. V. Averin, Phys. Rev.
Lett. 88, 050403(2002).

\bibitem{AC vor4} W. J. Elion, J. J. Wachters, L. L. Sohn, and J. E. Mooij, Phys. Rev. Lett. 71, 2311 (1993).

\bibitem{2002jso} E. V. Bezuglyi, A. S. Rozhavsky, I. D. Vagner, and P. Wyder,
Phys. Rev. B 66, 052508(2002).

\bibitem{2006jra} O. V. Dimitrova and M. V. Feigel¡¯man, ELECTRONIC PROPERTIES
OF SOLIDS, 102, 652(2006).

\bibitem{2007jqd} L. Dell'Anna, A. Zazunov, R. Egger, and T.
Martin, Phys. Rev. B, 75, 085305 (2007).

\bibitem{2008jra} Z. H. Yang, Y. H. Yang, J. Wang, and K. S. Chan, JOURNAL OF APPLIED PHYSICS 103, 103905
(2008).

\bibitem{2008jso} B. B\'{e}ri, J. H. Bardarson, and C. W. J. Beenakker,
Phys. Rev. B 77, 045311 (2008).

\bibitem{Andreev1} A. F. Andreev, Sov. Phys. JETP 19, 1228(1964).

\bibitem{RMPAndreev} Guy Deutscher, Rev. Mod. Phys 77, 109(2005).

\bibitem{Andreev2} A. F. Andreev, Sov. Phys. JETP 22, 455(1966).

\bibitem{Kulik} I. O. Kulik, Zh. Eksp. Teor. Fiz. 57, 1745(1969)[Sov. Phys. JETP 30,
944(1970)].

\bibitem{Bagwell} Philip F. Bagwell, Phys. Rev. B 46,12573(1992).

\bibitem{prl1991} C. W. J. Beenakker, Phys. Rev. Letters
67,3836(1991).

\bibitem{JoFET1} T. D. Clark, R.J. Prance, A. D. C. Grassie: J.
Appl. Phys. 51, 2736(1980).

\bibitem{JoFET2} H. Takayanagi, T. Kawakami: Phys. Rev. Lett. 54,
2449(1985).

\bibitem{JoFET3} Yong-Joo Doh, Jorden /a. van Dam, Aarnoud L. Roest,
Erik P. A. Bakkers, Leo P. Kouwenhoven, Silvano De Franceschi,
Science, 309, 272(2005).


\bibitem{Mario} M. F. Borunda, Xin Liu, Alexey A. Kovalev, Xiong-Jun
Liu, T. Jungwirth, L. W. Molenkamp, and Jairo Sinova.

\bibitem{semi-sup} T. Sch\"{a}pers, "Superconductor/Semiconductor Junctions", Springer Tracts in Moden Physics, 174, 2001

\bibitem{Jian-Bai Xia} Jian-Bai Xia, Phys. Rev. B, 45, 3593(1992).

\bibitem{rashbaring} F. E. Meijer, A. F. Morpurgo, and T. M.
Klapwijk, Phys. Rev. B 66, 033107(2002).


\bibitem{tunneling} Vinay Ambegaokar and Alexis Baratoff, Phys. Rev.
                    Lett. 10, 486(1963).

\bibitem{Zagoskin} Alexandre M. Zagoskin, "Quantum Theory of
Many-Body Systems: techniques and applications", New York:
Springer(1998).

\bibitem{ring} Diego Frustaglia and Klaus Richter, Phys. Rev. B 69,
               235310(2004).
\bibitem{Mahan} Gerald D. Mahan, "Many-Particle Physics" (The second
Edition), Plenum Press $\bullet$ New York and London.

\bibitem{Data} Supriyo Datta, "Electronic Transport in Mesoscopic
Systems", Cambridge University Press.

\bibitem{BTK} G. E. Blonder, M. Tinkham, T.M. Klapwijk, Phys. Rev. B 25,4515 (1982).

\bibitem{Anderson} E. Abrahams, P. W. Anderson, D. C. Licciardello, and T. V. Ramakrishnan,
Phys. Rev. Lett. 42, 673 (1979).

\bibitem{rf3} A. A. Golubov, M. Yu. Kupriyanov, E. Il'ichev, Rev. Mod. Phys 76, 411 (2004).

\bibitem{sup} Charles, P. Poole, Jr. "The Handbook of superconductivity"

\bibitem{pith} E. A. Demler, G. B. Arnold, and M. R. Beasley, Phys.
Rev. B 55, 15 174 (1997).

\bibitem{piex1} V. V. Ryazanov, A. V. Veretennikov, V. A. Oboznov, A. Yu.
Rusanov, V. A. Larkin, A. A. Golubov, and J. Aarts,
Physica C 341-348, 1613 (2000).

\bibitem{piex2} V. V. Ryazanov, V. A. Oboznov, A. Yu. Rusanov, A. V.
Veretennikov, A. A. Golubov, and J. Aarts, Phys. Rev.
Lett. 86, 2427(2001).

\bibitem{piex3} A. V. Veretennikov, V. V. Ryazanov, V. A. Oboznov, A. Yu.
Rusanov, V. A. Larkin, and J. Aarts, Physica B 284-288,
495 (2000).

\bibitem{1982book} Antonio Barone, Gianfranco Partern\`{o}, "Physics and Applications of the Josephson
Effect", A Wiley-Interscience publication, New York, 1982.

\bibitem{micro} A. Mart\'{i}n-Rodero, F. J. Carc\'{i}a-Vidal, and A. Levy Yeyati, Phys. Rev. Lett, 72, 554 (1994).

\bibitem{rf1} R. Rifkin,B.S. Deaver,Phys. Rev. B 13 (1976) 3894.

\bibitem{rf2} E. Il'ichev et al.,Rev. Sci. Instr. 72 (2001) 1882.


\bibitem{rf4} E. Il'ichev et al., Physica, C, 352 (2001) 141.

\bibitem{rf5} M. Grajcar, M. Ebel, E. Il'ichev, R. K\"{u}rsten, T. Matsuyama, U. Merkt, Physica C 372-376 (2002) 27-30.

\bibitem{exp1} A. M. Marsh, D. A. Williams, and H. Ahmed, Phys. Rev. B 50, 8118 (1994).

\bibitem{exp2} Yong-Joo Doh, et al. Science 309, 272 (2005).

\end{thebibliography}
\end{document}